\documentclass{PoS}

\title{The birth of radio millisecond pulsars and their high-energy signature}

\ShortTitle{State changes of MSPs}

\author{\speaker{P. H. T. Tam}\thanks{Now at Sun Yat-Sen University, China}, K. L. Li, A. K. H. Kong\\
        National Tsing Hua University, Taiwan\\
        E-mail: \email{phtam@phys.nthu.edu.tw}}

\author{J. Takata, G. C. K. Leung, K.S. Cheng\\
        The University of Hong Kong, Hong Kong}

\author{C. Y. Hui\\
        Chungnam National University, Korea}

\abstract{Millisecond pulsars (MSPs) are thought to born in low-mass X-ray binaries when the neutron star has gained enough angular momentum from the accreting materials of its companion star. It is generally believed that a radio MSP is born when the neutron star stops accreting and enters a rotation-powered state. Exactly what happens during the transition time was poorly understood until a year ago. In the past year, observations have revealed a few objects that not only switched from one state to the other (as predicted in the above picture), but also have swung between the two states within weeks to years. In this work, we present observations of two of these transition objects (PSR J1023+0038 and XSS J12270-4859) and a theoretical framework that tries to explain their high-energy radiation.}

\FullConference{10th INTEGRAL Workshop: "A Synergistic View of the High Energy Sky" - Integral2014,\\
		15-19 September 2014\\
		Annapolis, MD, USA }

\begin{document}

\section{Introduction}
The formation and evolution of millisecond pulsars (MSPs) is a major question in astrophysics. It is widely believed that neutron stars in low-mass X-ray binaries (LMXBs) can be spun up to millisecond periods by gaining angular momentum from the accreting materials of the companion star (e.g.~\cite{alpar82}). When the accretion stops, the neutron stars become radio millisecond pulsars powered by rotation (e.g.~\cite{campana98}). 

\begin{figure}
     \centering
     \includegraphics[width=.85\textwidth]{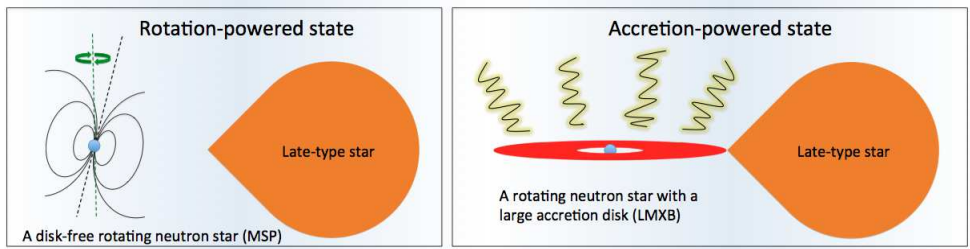}
     \caption{Two states of transitional MSPs}
     \label{graphic}
\end{figure}

At this stage, it was believed that the MSP changes from a LMXB to a rotation-powered state. Recent observations suggest that the binary can actually swing between the two states in time scales of weeks to years. A graphical representation of these two state is shown in Fig.~\ref{graphic}. In the following, we briefly describe the basic observational facts related to the identification of these two states of two iconic systems in the Galactic field: PSR J1023+0038 and XSS J12270-4859.

\section{PSR J1023+0038}
FIRST J102347.6+003841 was identified as a LMXB with an orbital period of 4.75 hours based on its X-ray and optical properties~\cite{thorstensen_armstrong05,homer06}. The source possessed an accretion disk between 2000 and 2001~\cite{wang09} but the disk has disappeared thereafter~\cite{archibald_sci_09}. Serendipitous observations in 2007 identified FIRST J102347.6+003841 as the radio PSR~J1023+0038 which has a spin period of 1.69~ms~\cite{archibald_sci_09}. Therefore, PSR~J1023+0038 is considered as a newly born MSP, representing the missing link of a rotation-powered MSP descended from a LMXB. PSR~J1023+0038 was also discovered as a $\gamma$-ray source in a dedicated analysis~\cite{tam10}. Although the origin of the $\gamma$-rays was ambiguous due to the lack of detection of modulation according to the spin nor orbital period, it was believed to be from the outer magnetosphere~\cite{takata10}. It was believed that the accretion disk of PSR~J1023+0038 would reappear once given a higher mass transfer rate, and the radio MSP would then be switched off~\cite{archibald_sci_09,wang09,tam10}. One very recent example is IGR J18245-2452 in the globular cluster M28 for which it has been switching between an accretion and a rotation-powered MSP~\cite{pappito13_m28}.

\begin{figure}
     \centering
     \includegraphics[width=.85\textwidth]{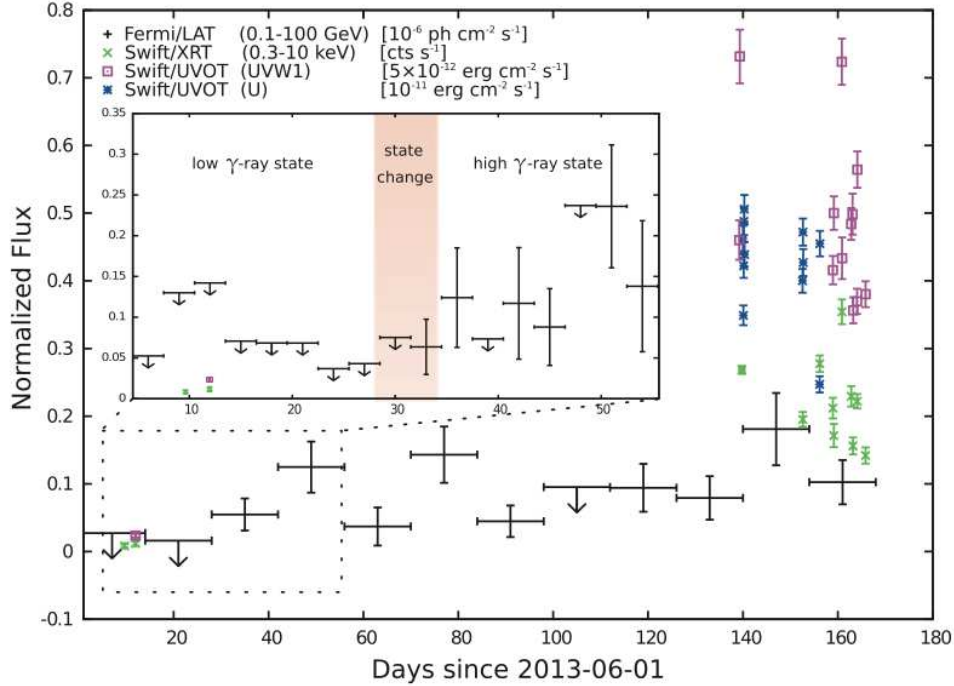}
     \caption{Multi-wavelength lightcurves of PSR~J1023+0038 from June 1 to November 13, 2013. The inset box indicates the evolution of the $\gamma$-ray emissions between June 6 and 24 July~\cite{takata14}. The state change (as inferred from $\gamma$-ray emission) occurred around July 2.}
     \label{state_change_1023}
\end{figure}

\begin{figure}
     \centering
     \includegraphics[width=.9\textwidth]{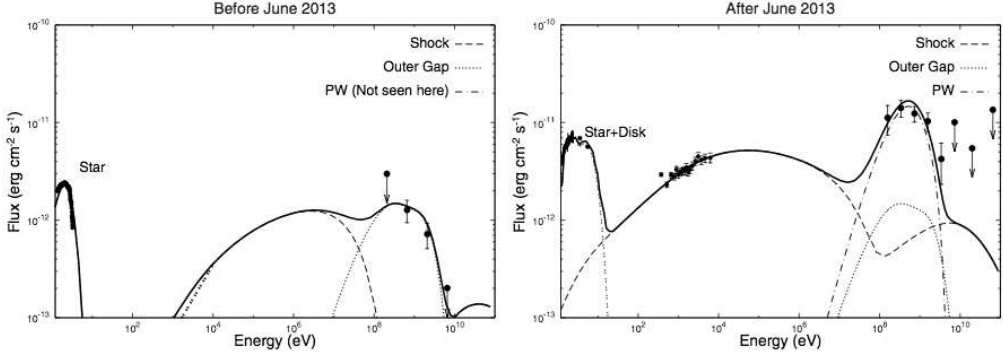}
     \caption{Modeling of the multi-wavelength spectra of PSR~J1023+0038 before and after the state change in 2013. The dashed, dotted, and dashed-dotted line represents the calculated spectra of the emissions from the shock, outer magnetospheric gap, and cold relativistic pulsar wind, respectively~\cite{takata14}.}
     \label{model_1023}
\end{figure}

Quite unexpectedly, the radio pulsation of PSR~J1023+0038 has disappeared since mid-June 2013~\cite{patruno14,stappers14}. Optical spectroscopy also showed strong double peaked $H_\alpha$ emission inferring that an accretion disk had been formed again~\cite{halpern13}. Moreover, the X-ray emission has increased by more than an order of magnitude from previous quiescent values and the UV emission has also brightened by 4 magnitudes~\cite{patruno14}. All these strongly indicate that PSR~J1023+0038 switched from a rotation-powered MSP to a LMXB with an accretion disk around mid-June 2013, in an opposite direction to what happened between 2002 and 2007, during which the system transited from a LMXB to a radio MSP. In addition, the $\gamma$-ray emission ($>$100~MeV) as seen with Fermi has brightened by a factor of five in early July (see Fig.~\ref{state_change_1023}; \cite{takata14}). It is somewhat unusual as we expect that the $\gamma$-rays should be turned off like radio during the accretion state (e.g., \cite{takata10}). In the model of~\cite{takata14}, X-rays are produced in the intra-binary shock while $\gamma$-rays are produced through inverse Compton up-scatterng off the enhanced ultra-violet photons emitted from the inner part of the accretion disk (see Fig.~\ref{model_1023}). A scenario is that the radio pulsar is still on even in the current high state but the radio pulses are blocked by the large amount of surrounding ionized material. The high rotation power of the pulsar interacting with companion star and the accretion disk formed again since June 2013 seems able to explain the optical, UV, and X-ray emission from the system~\cite{coti14}.

The state change can be seen in hard X-rays as well: NuSTAR/FPMA observed a flux increase in 3--79~keV by an order of magnitude in October 2013, as compared to June 2013. Orbital modulation is reported for the June 2013 data set, but not in October 2013~\cite{tendulkar14,li14}. Li et al.~\cite{li14} further discovered a $\sim$3130s modulation in NuSTAR data taken during the current LMXB state.

Continuous multiwavelength observations of the current high state of PSR~J1023+0038 will help us to understand the emission mechanisms as well as to probe its physical conditions, before it possibly return back to the rotation-powered MSP state in the near future.

\section{XSS J12270-4859}
XSS~J12270-4859 was the only LMXB possibly associated with a $\gamma$-ray source in the first Fermi/LAT catalog (1FGL~J1227.9-4852; \cite{martino10}); it also has a constant INTEGRAL flux~\cite{pappito14}. The source has a pulsar-like spectrum in the LAT energy range and is studied extensively in X-rays with Swift, RXTE, XMM-Newton, and Suzaku (e.g., \cite{saitou09,martino10,martino13}). However, its origin was not well understood given its peculiar multiwavelength properties and several possible radio counterparts, the brightest one being an AGN. Hill et al.~\cite{hill_11} first noted similarities of XSS~J12270-4859 with PSR~J1023+0038 but were not able to draw firm conclusion.

The nature of XSS~J12270-4859 became clearer after some follow-up investigations triggered by the 2013 state change of PSR~J1023+0038 that is described in the previous section. Optical and X-ray brightness was found to suddenly decrease some time between November 14, 2012 and December 21, 2012~\cite{bassa14,bogdanov14}. de Martino et al. \cite{martino14} reported a dramatic change in optical spectrum of this source, showing that the accretion disk that existed as late as March/April 2012 has disappeared in December 2013. These observations all point to the notion that XSS J12270-4859 is a similar system to PSR J1023+0038, but had undergone a state change from a LMXB to a rotation-powered state.

In analogous to PSR~J1023+0038, it was expected that radio pulsation from XSS~J12270-4859 should appear during the current optical/X-ray low (rotation-powered) state. The discovery of PSR~J1227-4853 confirms that the system now hosts a radio MSP~\cite{roy_atel,roy14}.

\begin{figure}
     \centering
     \includegraphics[width=.7\textwidth]{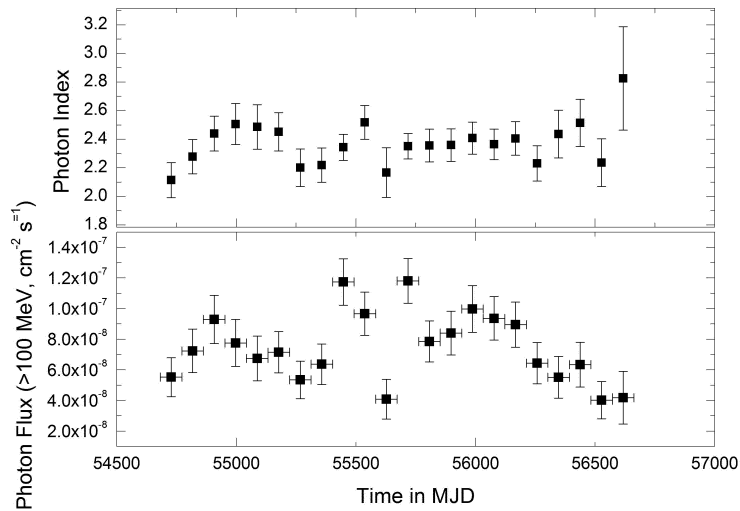}
     \caption{$\gamma$-ray results of XSS~J12270-4859 from August 2008 to February 2014, with a bin size of 90 days. \emph{Upper panel}: Power law is assumed and the evolution of the power-law index is shown. \emph{Lower panel}: $\gamma$-ray light curve at $>$100~MeV. The state change seen in optical and X-rays occurred between November 14 (MJD 56245) and December 21, 2012 (MJD 56282).}
     \label{gamma_lc}
\end{figure}

To further study the high-energy properties of the system, we constructed the $\gamma$-ray light curve from August 2008 to February 2014, as shown in Fig.~\ref{gamma_lc}, with a bin size of 90 days. Power law is assumed for XSS~J12270-4859 and the evolution of the power-law index is also shown. The time span of the light curve partly coincides with those reported in~\cite{hill_11,martino13}, who found no change in the $\gamma$-ray flux over their period of study. A major difference between their and our analysis is that aperture photometry was used in these previous works, and thus background photons including those from neighboring point sources may contaminate the results. Very recently and after this work had been presented in the workshop, Xing \& Wang~\cite{xing14} also studied the $\gamma$-ray behaviour of the system and their results are consistent with ours.

\begin{figure}
     \centering
     \includegraphics[width=.7\textwidth]{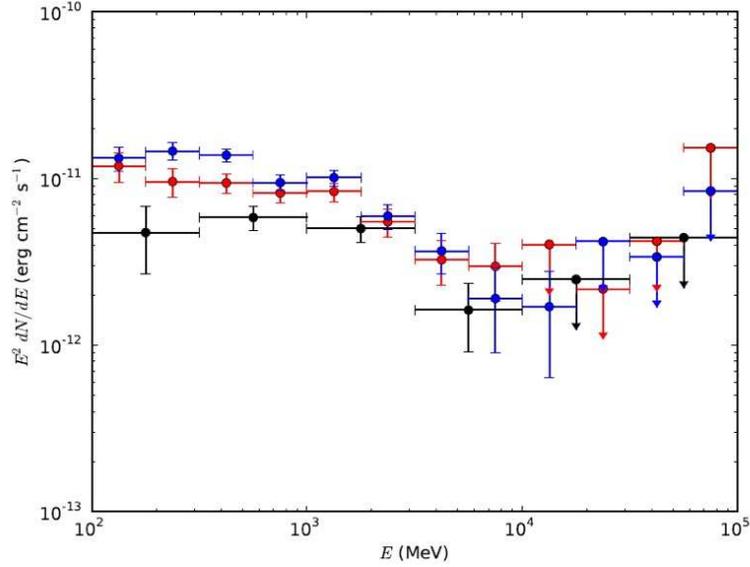}
     \caption{The spectrum of XSS~J12270-4859 for the phase I, II (before state change), and III (after state change) is shown in red, blue, and black, respectively.}
     \label{gamma_spec}
\end{figure}

The 5.5-year LAT light curve shows some fluctuation in flux that is unrelated to the state change occurred in end 2012. We further studied its spectrum by dividing the $\gamma$-ray data into three epochs: (I): MJD 54685 to MJD 55400, (II): MJD 55400 to MJD 56230, and (III): MJD 56230 to 56700. As shown in Fig.~\ref{gamma_spec}, there is no substantial change in the $\gamma$-ray spectrum between (I) and (II), while the spectrum in (III) differs significantly from the other two epochs. Therefore we conclude that the state change also occurred in $\gamma$-rays, but not as obvious in the total photon flux as in form of spectral change.

\begin{figure}
     \centering
     \includegraphics[width=.9\textwidth]{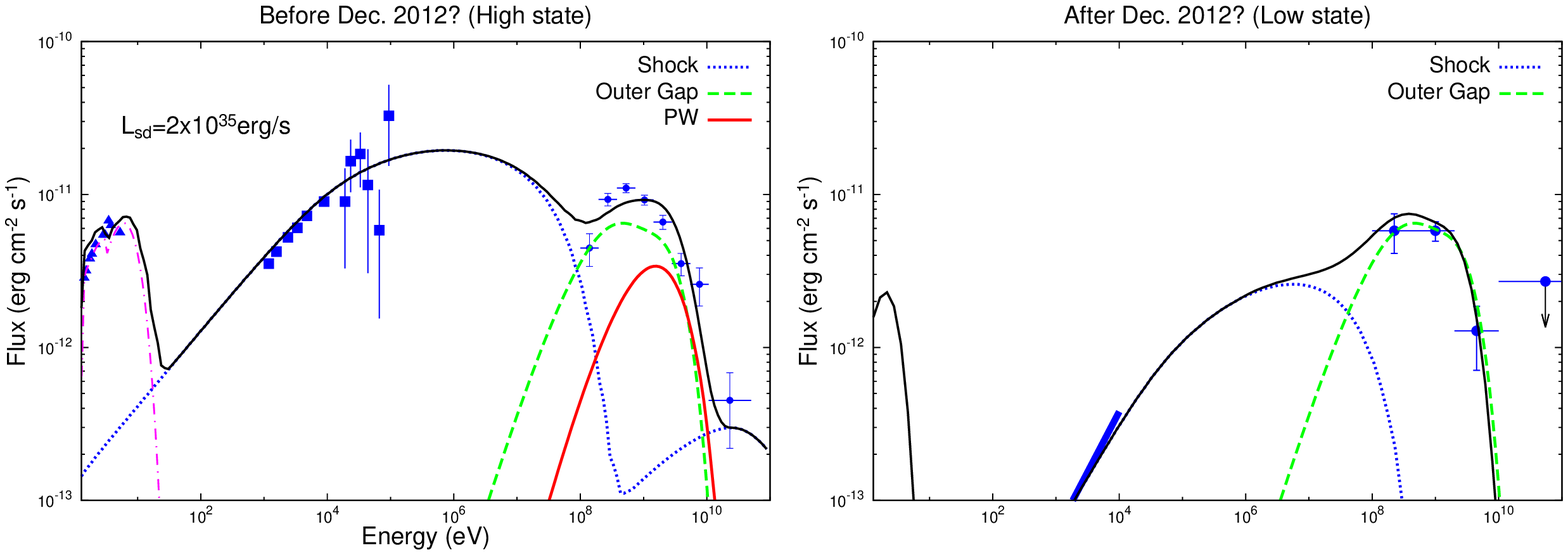}
     \caption{Modeling of the multi-wavelength spectra of XSS~J12270-4859 before and after the state change in end 2012. The blue dotted, green dashed, and red solid line represents the calculated spectra of the emissions from the shock, outer magnetospheric gap, and cold relativistic pulsar wind, respectively. A model similar to that presented in~\cite{takata14} is used.}
     \label{model_1227}
\end{figure}

We model the spectral energy distribtion from optical to high-energy $\gamma$-rays using the model originally for PSR~J1023+0038~\cite{takata14}. This is shown in Fig.~\ref{model_1227}. Hard X-ray (e.g., NuSTAR) or soft $\gamma$-ray observations during the current low state will be very crucial to probe the unexplored spectral region between 10~keV and 100~MeV. Continuous observations are also very important in constraining the emission regions.

\end{document}